\documentclass[prl,aps,twocolumn]{revtex4}
\usepackage{epsfig}

\begin{document}

% \draft
\title{Comment on ''Dynamic Scaling of Non-Euclidean Interfaces''}
\author{Joachim Krug}
\address{Institut f\"ur Theoretische Physik, Universit\"at zu K\"oln, Z\"ulpicher Str. 77, 50937 K\"oln, Germany}
\date{\today}

\vspace*{0.5cm}

% \pacs{81.10.Aj, 68.35.Ct, 02.40.Ky, 05.40.-a, 68.35.Fx}

\maketitle
% \begin{abstract}

% \end{abstract}

% \newpage

In a recent Letter Escudero \cite{Escudero} claims that the scaling properties
of curved growing interfaces are fundamentally different from those of their planar
counterparts. Specifically, based on an analysis of linear stochastic growth
equations in a radial geometry, he argues that the interface roughness
generically grows only logarithmically in cases where the planar interface
shows power law roughening. This claim is remarkable, because it contradicts
much of the work on kinetic roughening, where the equivalence of 
scaling properties between radial and planar growth geometries has been 
demonstrated in numerous simulations \cite{KS}, as well as recent rigorous
work on growth models in the Kardar-Parisi-Zhang (KPZ) universality class, in which
the dependence on growth geometry can be analyzed in great detail
\cite{Spohn}.

In fact I will argue here that Escudero's claim is erroneous. The analysis
of his calculation is aided by the fact the problem on which his argument is
based, the one-dimensional Edwards-Wilkinson (EW) equation in a radial geometry, has been solved
previously by Singha \cite{Singha}. We start from 
Escudero's equation (15) [referred to as (E15) in the following]
\begin{equation}
\label{fluc}
\partial_t \rho = D(Ft)^{-2} (\partial^2_\theta \rho + \rho) + (Ft)^{-1/2} \xi(\theta,t)
\end{equation}
for the radial surface fluctuation $\rho(\theta,t)$, which is solved  
 by decomposing $\rho(\theta,t)$ into radial Fourier modes $\rho_n(t)$.
Equation (\ref{fluc}) 
is identical to Singha's equation (7), apart from the term proportional to 
$\rho$, which derives from the curvature term 
$D/r$ in the radial EW-equation (E12).
The additional term can be easily transformed away and
does not affect the asymptotic results on time scales large compared to 
$D/F^2$. The equation (E19) for the two-point correlation $\langle \rho_n(t) 
\rho_m(t) \rangle$ of Fourier modes 
matches the corresponding expression in \cite{Singha}, with the
difference that Singha considers $\langle \rho_n(t) \rho_{-n}(t') \rangle$.
Escudero then takes the limit $t \to \infty$, finding that
$
\langle \rho_n(t) \rho_m(t) \rangle \sim (2 \pi F)^{-1} \delta_{n, -m}  \ln t.
$
This is inserted into the expression 
\begin{equation}
\label{C}
C(\theta,\theta',t) \equiv \langle \rho(\theta,t) \rho(\theta',t) \rangle = 
\sum_{n,m} e^{i(n \theta + m \theta')} \langle \rho_n(t) \rho_m(t) \rangle
\end{equation}
for the full correlation function of radial surface fluctuations, yielding the 
central result
\begin{equation}
\label{CC}
C(\theta,\theta',t) \sim \frac{\ln t}{2 \pi F} \delta(\theta - \theta'),
\end{equation}
from which logarithmic roughening in the one-dimensional radial EW equation is concluded. 

However, (\ref{CC}) in itself contains no information about the
surface roughness $w = \sqrt{C(\theta,\theta,t)}$.  
For $\theta \neq \theta'$ Eq.(\ref{CC}) states that
points on the interface separated by a finite angular difference become uncorrelated for
long times. This is a simple consequence of the standard scaling
theory of kinetic roughening \cite{KS,Krug}, in  
which the two-point correlation function takes on the scaling form
\begin{equation}
\label{Cscal}
C(\theta,\theta',t) = t^{2 \beta} {\cal{C}}(\ell/\xi) \approx t^{2 \beta} {\cal{C}}(\vert \theta
-\theta' \vert t^{1-1/z}),
\end{equation}
where $\ell \sim t \vert \theta - \theta' \vert$ is 
the distance between the angular points $\theta, \theta'$ measured
along the surface, 
and the correlation length $\xi \sim t^{1/z}$ with $z = 2$ for the EW equation.
For $t \to \infty$ and fixed $\vert \theta - \theta' \vert$, Eq.(\ref{Cscal}) reduces to 
\begin{equation}
\label{Cscal2}
\lim_{t \to \infty} C(\theta,\theta',t) \sim t^{2 \beta - 1 + 1/z} \delta(\theta - \theta').
\end{equation} 
For general linear growth equations the scaling relation \cite{Krug}
$2 \beta = 1 - d/z$ implies that the exponent of the prefactor in (\ref{Cscal2})
becomes $-(d-1)/z$, which vanishes for $d=1$, consistent with the
logarithmic behavior in (\ref{CC}). But clearly the logarithmic growth
of the prefactor in (\ref{CC}) does \textit{not} imply logarithmic
growth of $w$, in the sense of $\beta = 0$.  
The explicit calculation of the surface width in the radial geometry
requires the summation of the series (\ref{C}) at finite $t$. 
This was carried out by Singha, who finds that the surface roughness
grows as $w \sim \sqrt{t}$, exactly as in the planar case, but with a
different numerical prefactor.

Contrary to the claims in \cite{Escudero},
there is no need to revise the phenomenological scaling theory
of surface fluctuations \cite{KS,Krug},
which states that scaling exponents are robust with respect to 
growth geometry. 
In contrast, refined quantities such as universal amplitudes 
\cite{Krug91} and scaling functions do depend on the global boundary conditions
of the process in subtle ways 
that are currently being explored \cite{Spohn}.

\end{document}